\documentstyle[12pt,titlepage]{article}

\renewcommand{\theequation}{\thesection.\arabic{equation}}

\font\medio=cmr10 scaled \magstep2
\outer\def\beginsection#1\par{\medbreak\bigskip
      \message{#1}\leftline{\bf#1}\nobreak\medskip
\vskip-\parskip
      \noindent}

\def\laq{\raise 0.4ex\hbox{$<$}\kern -0.8em\lower 0.62
ex\hbox{$\sim$}}
\def\gaq{\raise 0.4ex\hbox{$>$}\kern -0.7em\lower 0.62
ex\hbox{$\sim$}}

\def\beq{\begin{equation}}
\def\eeq{\end{equation}}
\def\bea{\begin{eqnarray}}
\def\eea{\end{eqnarray}}

\def \pa {\partial}
\def \ra {\rightarrow}

\def \fb {\overline \phi}

\def \la {\lambda}
\def \ls {\lambda_s}

\def \b {\beta}
\def \a {\alpha}
\def \ap {\alpha^{\prime}}

\def \Ga {\Gamma}
\def \ga {\gamma}

\def \da {\delta}

\def \Om {\Omega}

\def \pfb {\Pi_{\fb}}
\def \pM {\Pi_{M}}
\def \pbe {\Pi_{\b}}

\begin{document}
\bibliographystyle {unsrt}

\titlepage
\begin{flushright}
CERN-TH/96-32 \\
IP-BBSR/96-9 \\
\end{flushright}
\vspace{10mm}
\begin{center}
{\bf GRACEFUL EXIT IN QUANTUM STRING COSMOLOGY}\\

\vspace{10mm}

M. Gasperini\footnote{Permanent address: 
{Dipartimento di Fisica Teorica, Via P. Giuria 1, 10125 Turin,
Italy.}} \\
{\em Theory Division, CERN, CH-1211 Geneva 23, Switzerland} \\
J. Maharana\footnote{Jawaharlal Nehru Fellow.}\\
{\em Institute of Physics, Bhubaneswar 751 005, India} \\
and\\
G. Veneziano \\
{\em Theory Division, CERN, CH-1211 Geneva 23, Switzerland} \\
\end{center}
\vspace{10mm}
\centerline{\medio  Abstract}

\noindent
We write an $O(d,d)$-covariant Wheeler-De Witt equation in the 
($d^2+1$)-dimensional minisuperspace of low-energy cosmological  
string backgrounds. We discuss explicit examples of transitions between two 
duality-related cosmological phases, and we find a finite 
quantum transition probability even when the two phases 
are classically separated by a curvature singularity. This quantum 
approach is completely free from operator ordering ambiguities 
as a consequence of the duality symmetries of the string 
effective action. 

\vspace{5mm}

\vfill
\begin{flushleft}
CERN-TH/96-32 \\
February 1996 
\end{flushleft}

\newpage

\renewcommand{\theequation}{1.\arabic{equation}}
\setcounter{equation}{0}
\section {Introduction}

The cosmological solutions of the low-energy string effective action, 
and their duality symmetry properties, have recently motivated the study 
of an inflationary scenario \cite{1}-\cite{3} in which the universe 
evolves from the string perturbative vacuum to the high curvature, 
strong coupling regime. One of the main problems, in this context, is 
the ``graceful" exit from such an accelerated (also called pre-big-bang 
\cite{1}) phase to the decelerated, decreasing curvature phase typical 
of the standard (post-big-bang) cosmological evolution. Both phases are 
present in the exact solutions of the tree-level, lowest order in $\ap$, 
string effective action \cite{2,3}. However, for vanishing torsion and 
dilaton potential, these phases correspond to different 
(duality-related) branches of the 
solution, defined over disconnected ranges of the time parameter and 
separated by a singularity of the curvature and of the coupling.

The graceful exit problem would be solved, at the classical level, by an 
exact cosmological solution connecting smoothly the two branches, and 
thus describing a continuous evolution from accelerated to decelerated 
expansion. Unfortunately, confirming a previous conjecture \cite{4}, it 
has been rigorously proved \cite{5} that such a change of branch cannot 
be simply catalyzed by any (realistic) dilaton potential, 
if we limit ourselves to lowest order in the $\ap$ expansion of the 
string effective action. Such a no-go theorem has been recently 
extended, for spatially homogeneous and isotropic manifolds, to the case 
of non-vanishing torsion background \cite{6} and non-vanishing spatial 
curvature \cite{7}. 

A possible way of incorporating branch-changing in the pre-big-bang 
scenario is thus to resort to the conformal field theory approach 
\cite{8}, where all higher orders in $\ap$ are taken into account. On the 
other hand, in such a ``stringy" regime dominated by higher-derivative 
terms in  the effective action, the curvature is expected to approach 
the Planck scale and thus the quantum gravity regime. This suggests a 
quantum approach to the graceful exit problem (quantum cosmology methods 
in a string theory context were previously introduced also in 
\cite{9,10}).

By applying the Wheeler-De Witt (WDW) equation to the gravi-dilaton 
system, we show in this paper that the transition from a pre-big-bang 
to a post-big-bang classical solution corresponds to a reflection of the 
wave function in minisuperspace. A classical configuration describing 
branch-changing gives a reflection coefficient $R=1$. The reflection 
probability is in general non-vanishing, however, even if the given 
classical background forbids branch-changing. This is the main result of 
this paper, which may allow a systematic classification of the initial 
conditions compatible with the present state of our Universe, even 
ignoring kinematical details during the quantum transition era. 

The paper is organized as follows. In Section 2 we derive from the low 
energy string effective action the WDW equation for homogeneous, 
spatially flat cosmological backgrounds, including a non-trivial 
antisymmetric (torsion) tensor. We show, in particular, that the 
operator ordering problem is trivially solved because of the $O(d,d)$ 
covariance of the kinetic part of the Hamiltonian, which implies a 
globally flat minisuperspace metric. In Section 3 we study the free wave 
equation, and we identify the left and right moving modes in superspace 
with the two branches of the classical vacuum solution. The reader not 
interested in technical complications due to the presence of a 
non-trivial torsion background can move directly to Section 4, where we 
give two self-contained examples of ``quantum" branch changing, in the  
two-dimensional minisuperspace parameterized  by the  
dilaton field and by the (isotropic) metric scale factor. While the 
first example has a classic analogue, the second corresponds to a background 
configuration in which branch changing is classically forbidden. A brief 
summary, and our concluding remarks, are finally presented in Section 5.

\vskip 2 cm
\renewcommand{\theequation}{2.\arabic{equation}}
\setcounter{equation}{0}
\section{O(d,d)-covariant Wheeler-De Witt equation}

At low energy, the tree-level, $(d+1)$-dimensional (super)string 
effective action can be written as \cite{12}
\beq
S=-{1\over 2\la_s^{d-1}}\int d^{d+1}x \sqrt{|g|}e^{-\phi}\left(R+
\pa_\mu\phi\pa^\mu\phi-{1\over 12}H_{\mu\nu\a}H^{\mu\nu\a}+V\right).
\label{21}
\eeq
Here $\phi$ is the dilaton field, $H_{\mu\nu\a}$ is the field strength 
of the two-index antisymmetric torsion tensor $B_{\mu\nu}=-B_{\nu\mu}$, 
and $\la_s\equiv (\ap)^{1/2}$ is the fundamental string length parameter 
governing the high derivative expansion of the action. Note that we have 
included a possible dilaton potential $V$,  
and we have chosen to work in the String (or Brans-Dicke frame), whose 
metric coincides with the sigma-model metric to which strings are 
directly coupled. The more conventional choice of the Einstein frame 
leads to an equivalent description of the same cosmological scenario 
\cite{3,13}, but it is less convenient for exploiting the duality 
symmetries of the underlying theory.

We shall consider, in this paper, homogeneous backgrounds with $d$ 
Abelian isometries, for which a synchronous frame exists where 
$g_{00}=1$, $g_{0i}=0=B_{0i}$, and the fields are independent of all 
space-like coordinates $x^i$ ($i,j=1,..,d$). We also assume spatial 
sections of finite volume, $(\int d^dx\sqrt{|g|})_{t={\rm const}}<\infty$. For 
such backgrounds the action can be rewritten as \cite{14}
\beq
S=-{\ls\over 2}\int dt e^{-\fb}\left[(\dot{\fb})^2+{1\over 8}{\rm Tr}
~\dot M(M^{-1})\dot{}+V\right]
\label{22}
\eeq
where a dot denotes differentiation with respect to the cosmic time $t$, 
and we have chosen to express length and energies in string units 
through $\ls$. Here $\fb$ is the `` shifted" dilaton field,
\beq
\fb=\phi-\ln|{\rm det}~ g_{\mu\nu}|^{1/2}
\label{23}
\eeq
(we have absorbed into $\phi$ the constant shift $-\ln(\ls^{-d}\int 
d^dx)$ required to secure the scalar behaviour of $\fb$ under coordinate 
reparametrization). Finally, $M$ is the $2d\times 2d$ matrix
\beq
M=\pmatrix{G^{-1} & -G^{-1}B \cr
BG^{-1} & G-BG^{-1}B \cr}
\label{24}
\eeq
where $G$ and $B$ are, respectively, matrix representations of the spatial 
part of the metric ($g_{ij}$) and of the antisymmetric tensor ($B_{ij}$). 

For constant $V$, the whole action (\ref{22}) is invariant 
under global $O(d,d)$ transformations \cite{14}
\beq
\fb \ra \fb , ~~~~~~~~~~ M\ra \Om^T M \Om
\label{25}
\eeq
where
\beq
\Om^T\eta \Om =\eta, ~~~~~~~~~~ \eta =
\pmatrix{0 & I \cr I & 0 \cr} .
\label{26}
\eeq
In addition, $M$ satisfies 
\beq
M\eta M = \eta.
\label{27}
\eeq
This $O(d,d)$ symmetry is preserved in the presence of bulk string 
matter \cite{15} satisfying the string equations of motion, and it 
reduces to the scale factor duality symmetry \cite{16,17} (for 
torsionless, diagonal metric backgrounds) in the particular case in 
which 
we restrict $\Om$ to $\eta$ in eq. (\ref{25}). 

By using as time parameter $\tau$, with 
$dt=e^{-\fb}d\tau$, the action (2.2) leads to the Lagrangian
(a prime denotes differentiation with respect to $\tau$)
\beq
L(\tau)=-{\ls \over 2}\left[(\fb')^2+{1\over 8}{\rm Tr}~ M'(M^{-1})'+ 
e^{-2\fb}V\right]
\label{28}
\eeq
whose corresponding Hamiltonian is 
\beq
H= -{1\over 2\ls}\pfb^2+{4\over \ls}{\rm Tr}~(M~\Pi_M M~\Pi_M)
+{\ls \over 2}V e^{-2\fb}
\label{29}
\eeq
where $\pfb$ and $\Pi_M$ are the (dimensionless) canonical momenta
\beq
\pfb ={\da L \over \da \fb '}= -\ls \fb' ,~~~~~~~~~~~ 
\pM={\da L \over \da M '}= {\ls \over 8}M^{-1}M'M^{-1} .
\label{210}
\eeq

The variation of the action (\ref{21}) with respect to the ``lapse" 
function, $\sqrt{g_{00}}$, provides the canonical constraint $H=0$. The 
WDW equation \cite{18}, implementing in superspace such a constraint 
through the differential representation $\pfb =\pm i \da/\da\fb$, 
$\pM =\pm i \da /\da M$, would seem to be affected (as usual) by 
problems of quantum ordering, as $[M,\pM]\not= 0$. In our context, 
however, the problems actually disappear because our 
$(d^2+1)$-dimensional minisuperspace is globally flat, as a consequence 
of the $O(d,d)$ symmetry. Indeed, by using the $O(d,d)$ property 
(\ref{27}), we can always rewrite the $M$-dependent part of the kinetic 
operator as
\beq
{1\over 16}{\rm Tr}~ M'(M^{-1})'=
{1\over 16}{\rm Tr}~ (M'\eta)^2.
\label {211}
\eeq
The corresponding Hamiltonian
\beq
H= -{1\over 2\ls}\pfb^2-{4\over \ls}{\rm Tr}~(\eta~\Pi_M~ \eta~\Pi_M)
+{\ls \over 2}V e^{-2\fb}
\label{212}
\eeq
has a flat metric in momentum space, and leads to a WDW equation
\beq
\left[{\da^2 \over \da \fb^2}+ 8{\rm Tr}~\left(\eta {\da \over \da M}
\eta {\da \over \da M}\right) +\ls^2 V e^{-2\fb}~ \right]\Psi(\fb, M)=0,
\label{213}
\eeq
which is manifestly free from problems of quantum ordering.

If we introduce curvilinear coordinates in minisuperspace, adopting for 
instance the parametrization of eq. (\ref{29}), the ordering imposed by 
the $O(d,d)$ symmetry is equivalent to the general covariance of the 
Laplacian operator, as can be easily checked for the simple isotropic 
case $B=0$, $G_{ij}=-a^2\da_{ij}$. In that case the kinetic part of the 
Hamiltonian (\ref{212}) is represented as
\beq
\ls H_{\rm Kin}=- {1\over 2} \pfb^2 -4 {\rm Tr}~ (\eta \pM)^2
\equiv {1\over 2} \pa^2_{\fb}- 2 d(a\pa_a +a^2 \pa^2_a).
\label {214}
\eeq
The parametrization of eq. (\ref{29}), on the other hand, corresponds 
to the metric
\beq
\ga_{\mu\nu}={\rm diag}~ \left(-2,~ {a^4\over 4d}, 
~{1\over 4da^4}~\right),
~~~~~~~~~ \mu, \nu =1,2,3 ~~~,
\label{215}
\eeq
in the three-dimensional space spanned by the differential operators
\beq
\Pi_\mu= i\pa_\mu= i\left(~{\pa \over \pa \fb},~ {\pa\over \pa a^{-2}}, 
~{\pa\over \pa a^{2}}~\right) 
\label{216}
\eeq
and the covariant Laplacian gives
\beq
-\nabla_\mu \nabla^\mu \equiv -{1\over \sqrt{|\ga|}}
\pa_\mu \left (\sqrt{|\ga|}\ga^{\mu\nu}\pa_\nu \right)=
{1\over 2} \pa^2_{\fb}- 2 d(a\pa_a +a^2 \pa^2_a)
\equiv \ls H_{\rm Kin}.
\label{217}
\eeq
The ordering fixed by the scale factor duality symmetry of the classical 
Hamiltonian is thus the same as that imposed by the requirement of 
general reparametrization invariance in minisuperspace (note that in our 
case there is no possible contribution to the ordered Hamiltonian from 
the scalar curvature of superspace \cite{19}, as the metric is globally 
flat).

\vskip 2 cm

\renewcommand{\theequation}{3.\arabic{equation}}
\setcounter{equation}{0}
\section{Branch changing as wave reflection}

We shall apply, in this paper, the WDW equation (2.13) to study the 
probability of transition from a given initial background configuration 
of the pre-big-bang type, to a final configuration typical of the 
standard cosmological scenario. This amounts to solving  eq. (\ref{213}) 
for a given value of the dilaton potential $V(\phi)$, with appropriate 
boundary conditions. The systematic study of the transition probability 
for a ``realistic" (supersymmetry breaking) non-perturbative dilaton 
potential is postponed to a future work. The main goal of this paper is 
to show that in a quantum cosmology context it is possible to tunnel 
from one branch to another of the low energy classical solutions, even 
if the two branches are not smoothly connected, but they are separated 
by curvature singularities and unphysical regions of finite size.

To this aim we shall work in the simplifying hypothesis of $O(d,d)$ 
symmetry of the whole (in general non-local) action, assuming 
$V=V(\fb)$, because in that case the WDW equation (\ref{213}) can be 
separated by setting
\beq
\Psi(\fb,M)= \chi_A(M)\psi_A(\fb)
\label{31}
\eeq
where
\beq
(M\pM)\chi_A\equiv i M {\da \over \da M}\chi_A
 =-\left({1\over 8}A\eta\right) \chi_A
\label{32}
\eeq
and
\beq
\left[{\da^2 \over \da \fb^2} +{1\over 8}{\rm Tr}~ (A\eta)^2 +\ls^2
V(\fb)e^{-2\fb} \right] \psi_A(\fb)=0.
\label{33}
\eeq
We have used here the momentum conservation $[M \pM,H]=0$, in agreement with 
the classical equations of motion obtained from the Lagrangian 
(\ref{28}), which imply \cite{14,15}
\beq
M\pM = -{\ls \over 8}M \eta M' \eta = -{1\over 8} A \eta
\label{34}
\eeq
where $A$ is a constant $2d \times 2d$ matrix satisfying
\beq
M \eta A+ A\eta M =0.
\label{35}
\eeq
If we consider, in particular, the ``free" wave equation ($V=0$), 
eq. (\ref{33}) is easily solved by a linear superposition of left and 
right moving waves,
\beq
\psi^{\pm}_A(\fb)= \exp \left\{\pm {i\over 2}\fb \left[{1\over 2} 
{\rm Tr}~ (A\eta)^2\right]^{1/2}\right\}.
\label{36}
\eeq

For simplicity, we shall restrict our subsequent discussion to a 
diagonal, Bianchi I type vacuum background with 
$G_{ij}=-a^2_i(t)\da_{ij}$ and $B=0$, corresponding to \cite{14}
\beq
 A= \pmatrix{0 & -A_d \cr A_d & 0 \cr} ,~~~~~~~ 
(A_d)_{ij}=c_i\da_{ij}~~~,
\label{37}
\eeq
where $c_i$ are arbitrary constants. 
The metric-dependent part of the wave function 
becomes in this case 
\beq
\chi_A(a_j)= N \exp\left\{-{i\over 2}\sum_j c_j \ln a_j \right\} ,
\label{38}
\eeq
as one can check after realizing that 
the operator $M\pM$ contains in this case only $d$ independent 
variables. By defining 
\beq
\a_j = c_j \left(\sum_jc_j^2\right)^{-1/2}\equiv c_j \left[{1\over 2}
{\rm Tr}~(A\eta)^2\right]^{-1/2}, ~~~~~~~~~~~
\sum_j \a_j^2=1 
\label{39}
\eeq
the solutions of the WDW equation can finally be written in the form
\beq
\Psi_A^{(\pm)}(\fb,M)= \chi_A(M)\psi_A^{\pm}(\fb)=
N_\pm \exp\left\{-{i\over 2}\left[ {\rm Tr}~(A_d)^2\right]^{1/2}
\left(\sum_j 
\a_j \ln a_j \mp \fb~\right)\right\},
\label{310}
\eeq
where $N_{\pm}$ is an overall normalization coefficient.

For fixed $\a_j$, namely for a given eigenstate of $M\pM$, the left and 
right moving modes $\Psi^{(\pm)}$ correspond to different branches of the 
exact solution of the vacuum string cosmology equations 
\cite{2,3,14,16,17,19a}
\beq
a_j=a_{j0}|t/\ls|^{\pm \a_j} , ~~~~~~~ \sum_j\a_j^2=1 ,~~~~~~~~
\fb - \phi_0 = -\ln |t/\ls| = \mp \sum_j \a_j \ln a_j ,
\label {311}
\eeq
where $ a_{i0}$ and $\phi_0$ are integration constants. What is 
important, for our purpose, is that if we apply the momentum operator 
$\pfb= i \pa /\pa \fb$ to the right moving wave $\Psi_A^{(+)}$ 
(the opposite sign with respect to the standard convention is due to the 
definition of $\pfb$, eq. (\ref{210})), we reproduce the canonical 
momentum 
\beq
\pfb = -\ls \dot{\fb} e^{-\fb}= -e^{-\phi_0} <0
\label{312}
\eeq
of a classical configuration corresponding to an 
accelerated, expanding background, with growing curvature 
and dilaton coupling:
\beq
a_i \sim (-t)^{-\a_i},~~~~ t<0, ~~~~ \a_i >0, ~~~~
\fb - \phi_0 =\sum_j \a_j\ln a_j, ~~~~ \dot{\fb} >0 .
\label{313}
\eeq
By applying $\pfb$ to the left mover $\Psi^{(-)}_A$, labelled by the same 
eigenvalue $A$, we find instead a configuration with the opposite 
canonical momentum,
\beq
\pfb = -\ls \dot{\fb} e^{-\fb}= e^{\phi_0} >0
\label{314}
\eeq
corresponding again to an expanding branch of the same classical 
solution, but decelerated and with decreasing curvature: 
\beq
a_i \sim t^{\a_i},~~~~ t>0, ~~~~ \a_i >0, ~~~~
\fb - \phi_0 =-\sum_j \a_j\ln a_j, ~~~~ \dot{\fb} <0 .
\label{315}
\eeq
A branch transition of the type required to solve the graceful exit 
problem, involving a scale factor duality transformation and time 
reversal \cite{1}, namely 
$a(t)\ra a^{-1}(-t)$, is thus equivalent in this context to the spatial 
reflection of the WDW wave function in minisuperspace, $\Psi_A^{(+)}\ra 
\Psi_A^{(-)}$.

\vskip 2 cm

\renewcommand{\theequation}{4.\arabic{equation}}
\setcounter{equation}{0}
\section{Two simple examples}

As no reflection is possible for free waves, let us introduce an 
appropriate dilaton potential, considering for simplicity a $d=3$ 
isotropic background, and setting 
\beq
B=0,~~~~~~~~~~~~~~~ a(t)= \exp\left[\b(t)/\sqrt 3\right] .
\label{41}
\eeq
The lowest order gravi-dilaton effective action, 
\beq
S=-{1\over 2\ls}\int d^4x 
\sqrt{-g}e^{-\phi}\left(R+\pa_\mu\phi\pa^\mu\phi +V\right) ,
\label{42}
\eeq
after integrating by parts, and using as before the convenient time 
parameterization $dt=d\tau e^{-\fb}$, reduces to (in the gauge 
$g_{00}=1$):
\beq 
S=-{\ls\over 2}\int d\tau \left(~\fb^{\prime 2}-\b^{\prime 2}+Ve^{-2\fb} 
~\right)
\label{43}
\eeq
where
\beq
\fb= \phi -\ln\int (d^3x/\ls^3) - \sqrt 3 \b .
\label{44}
\eeq
The corresponding Hamiltonian
\beq
H= {1\over 2\ls}\left(~\pbe^2-\pfb^2+\ls^2V e^{-2\fb}~\right),
~~~~~~\pbe= \ls \b', ~~~~~ \pfb= -\ls \fb'
\label{45}
\eeq
is a particular case of eq. (\ref{212}), for the torsionless isotropic 
background considered here. The WDW equation then takes the general form 
of a two-dimensional Schr\"odinger-like equation in the plane ($\fb, \b$):
\beq
\left[~\pa^2_{\fb} - \pa^2_\b +\ls^2 V(\fb, 
\b)e^{-2\fb}~\right]\Psi(\fb,\b)=0 .
\label{46}
\eeq

As anticipated in the previous section, we shall assume in this paper 
$V=V(\fb)$, in order to separate variables. Let us discuss first  
the particular case
\beq
V(\fb)= -V_0e^{4\fb} , ~~~~~~~V_0={\rm const}, ~~~~~~~ V_0>0
\label{47}
\eeq
as a toy example of classical gravi-dilaton configuration allowing 
branch changing. A negative non-local dilaton potential, $V(\fb)<0$, 
although hard to motivate in a realistic superstring theory context, is 
indeed the only case in which exact analytical solutions are known 
\cite{1,20} connecting smoothly the pre- to the post-big-bang regime.

With the above potential, the classical equations of motion following 
from the Hamiltonian (\ref{45}) imply momentum conservation along the 
$\b$ axis,
\beq
\pbe=\ls \b' = \ls \dot \b e^{-\fb}=k = {\rm const}
\label{48}
\eeq
and are solved exactly by
\beq
\fb= -{1\over 2}\ln \left ({\ls^2 V_0\over k^2}+{k^2t^2\over 
\ls^2}\right) , ~~~~~~~ 
a= a_0\left[{k^2t\over 
\ls^2 \sqrt{V_0}}+\left(1+{k^4t^2\over\ls^4V_0}\right)^{1/2}
\right]^{1/\sqrt3}
\label{49}
\eeq
($a_0$ is a dimensionless integration constant). This is a regular 
``self-dual" solution, $a(t)/a_0=a_0/a(-t)$, characterized by a 
bell-like shape of the curvature scale and of the coupling $e^{\fb}$. It 
describes a background that evolves from an initial state of 
accelerated expansion and increasing curvature, 
\bea
t\ra -\infty,~~~~ &a& \sim(-t)^{-1/\sqrt3}, ~~~~ \fb \sim -\ln (-t)=
\sqrt 3 \ln a = \b \nonumber \\
&\dot a&>0, ~~~~ \ddot a >0, ~~~~ \dot H >0
\label{410}
\eea
to a final state of decelerated expansion, decreasing curvature, 
\bea
t\ra +\infty,~~~~ &a& \sim t^{1/\sqrt3}, ~~~~ \fb \sim -\ln (-t)=
-\sqrt 3 \ln a= -\b \nonumber \\
&\dot a&>0, ~~~~ \ddot a <0, ~~~~ \dot H <0 .
\label{411}
\eea

For the background generated by the potential (\ref{47}), the WDW 
equation (\ref{46}) can easily be separated by putting 
$\Psi(\fb,\b)$ = 
$e^{-ik\b}\psi_k(\fb)$, where $k$ belongs to the continuous eigenvalue 
spectrum of $\pbe$,
\beq
\pbe\Psi_k= i \pa_\b\Psi_k=k\Psi_k, ~~~~~~~[\pbe,H]=0
\label{412}
\eeq
and $\psi_k$ satisfies 
\beq
\left(~\pa^2_{\fb} +k^2- \ls^2V_0e^{2\fb}~\right)\psi_k(\fb)=0
\label{413}
\eeq
(note the role of time-like coordinate assigned to $\b$, monotonically 
ranging from $-\infty$ to $+\infty$). The general solution for $\psi_k$ 
is then a linear combination of modified Bessel functions $K_\nu(z)$, 
$I_\nu(z)$ \cite{21}, of complex index $\nu=ik$ and argument $z=\ls 
\sqrt{V_0}e^{\fb}$. We impose the regularity condition \cite{22} 
$|\Psi_k|<0$, corresponding to a vanishing wave function 
in the ``impenetrable" region of infinite effective potential, 
$\psi_k(\fb)\ra 0$ for $\fb \ra +\infty$. This condition uniquely selects 
(modulo a normalization factor) the WDW solution as 
\beq
\Psi_k (\fb,\b)= N K_{ik}(\ls\sqrt{V_0}e^{\fb}~)~ e^{-ik\b}.
\label{414}
\eeq

For $\fb \ra -\infty$, i.e. in the low energy regime, this solution 
contains asymptotically left and right moving waves, as \cite{21}
\bea
\lim_{\fb \ra -\infty} \Psi_k(\fb,\b) &=& 
-{N \pi\over 2 \sin(ik\pi)}\left[
\left(\ls\sqrt{V_0}\over 2\right)^{ik}{e^{-ik(\b-\fb)}\over\Ga(1+ik)}-
\left(\ls\sqrt{V_0}\over 2\right)^{-ik}{e^{-ik(\b+\fb)}\over\Ga(1-ik)}
\right]= \nonumber \\
&=&\Psi_k^{(+)}+\Psi_k^{(-)}
\label{415}
\eea
As discussed in the previous section, the right movers represent the 
accelerated, negative time branch (\ref{410}),  
with $\b=\fb$, the left movers the decelerated, positive time branch 
(\ref{411}), with $\b=-\fb$. The reflection coefficient, $R_k=
|\Psi_k^{(-)}|^2/ |\Psi_k^{(+)}|^2$, measures the probability of transition
between the two branches of the low energy classical solution. In the 
low energy limit, according to eq. (\ref{415}), $R_k\ra 1$ for all $k$, 
as expected because we have considered an example in which the two 
branches are smoothly connected already at the classical level.

Consider now an example with the opposite potential,
\beq
V(\fb)= V_0e^{4\fb} , ~~~~~~~V_0={\rm const}, ~~~~~~~ V_0>0 . 
\label{416}
\eeq
In this case branch changing is classically forbidden. Indeed, the 
momentum conservation (\ref{48}) is still valid, but the classical 
solution becomes
\beq
\fb= -{1\over 2}\ln \left ({k^2t^2\over
\ls^2} - {\ls^2 V_0\over k^2}\right) , ~~~~~~
a= a_0\left|{k^2t\over
\ls^2\sqrt{V_0}}+\left({k^4t^2\over\ls^4V_0}-1\right)^{1/2}
\right|^{1/\sqrt 3} . 
\label{417}
\eeq
The low energy (large time limit) branches (\ref{410}) and (\ref{411}) 
still exist, but they are now separated by an unphysical region, of extension 
$|t|<\ls^2\sqrt{V_0}/k^2$, where the dominant energy condition is 
violated, and the expansion rate ($H$) and the dilaton coupling 
($e^{\fb}$) become imaginary. A curvature singularity is present at both 
ends of such a region, where the branches (\ref{410}) and (\ref{411}) 
respectively end and start.

Nevertheless, the quantum probability of transition between the two 
branches is non-vanishing. In fact, the solution of the WDW equation 
can be factorized as before, with the difference that eq. (\ref{413}) is 
replaced by
\beq
\left(~\pa^2_{\fb} +k^2+ \ls^2V_0e^{2\fb}~\right)\psi_k(\fb)=0 . 
\label{418}
\eeq
The general solution for $\psi_k$ can now be written as a linear 
combination of first and second kind Hankel functions \cite{21}, 
$H_\nu^{(1)}(z)$ and $H_\nu^{(2)}(z)$. By assuming for the Universe an 
initial pre-big-bang configuration, we impose that in the high-curvature 
limit $z \ra \infty$ there are only right moving waves ($\dot{\fb} >0$, 
$\pfb <0$) approaching the singularity. This condition exactly coincides 
with the boundary conditions allowing 
tunnelling through classically forbidden regions of superspace \cite{22} 
(which select only outgoing waves at the superspace boundary, where 
classical trajectories can end but not begin), and fixes the wave 
function as
\beq
\Psi_k (\fb,\b)= N H_{ik}^{(1)}(\ls\sqrt{V_0}e^{\fb}~)~ e^{-ik\b}.
\label{419}
\eeq
Asymptotically, in the low curvature, perturbative regime $\fb \ra 
-\infty$, we then have 
\bea
\lim_{\fb \ra -\infty} \Psi_k(\fb,\b) &=&
{iN  \csc(ik\pi)}\left[e^{k\pi}
\left(\ls \sqrt{V_0}\over 2\right)^{ik}{e^{-ik(\b- \fb)}\over\Ga(1+ik)}-
\left(\ls\sqrt{V_0}\over 2\right)^{-ik}{e^{-ik(\b+\fb)}\over\Ga(1-ik)}
\right]= \nonumber \\
&=&\Psi_k^{(+)}+\Psi_k^{(-)}
\label{420}
\eea
and the relative amplitude of left and right modes defines the 
probability
\beq
R_k={|\Psi_k^{(-)}|^2\over |\Psi_k^{(+)}|^2}= e^{-2\pi k}
\label{421}
\eeq
for transitions from the classical trajectory with $\b=\fb$ to the 
duality-related one, $\b=-\fb$.

By recalling the definition of $k$ (eq. (\ref{48})) and of $\fb$,
\beq
k={\sqrt3\over \ls^2}\int d^3x\sqrt{-g} e^{-\phi}H= {\rm const}
\label {422}
\eeq
we can eventually express the above transition probability as
\beq
R(g_s,a_s)=\exp\left\{-{\sqrt{12}\pi \over g_s^2}{\Om(a_s)\over 
\ls^3}\right\}. 
\label{423}
\eeq
Here $\Om (a_s)$ and $g_s=e^{\phi_s/2}$ are, respectively, the values of 
the proper spatial volume and of the coupling, at the time $t=t_s$ at 
which $H=\ls^{-1}$. For values of the coupling  
$g_s \sim 1$ the probability (\ref{421}) is of order 1 for the 
formation of ``bubbles" of unit proper size (or smaller) in string units 
at $t=t_s$. 

The above example is not ``realistic", in the sense that it does not 
describe the formation of a radiation-dominated or matter-dominated 
Universe similar to the one we live in today (we postpone the discussion 
of a more realistic scenario to a forthcoming paper). Nevertheless, it 
is an example of how the Universe can emerge from the inflationary 
phase in the right branch corresponding to decelerated expansion, and it 
is quite interesting that the probability of such a process is peaked in 
the strong coupling regime, with a 
typical instanton-like behaviour $R\sim \exp(-g^{-2})$. 

We note, finally, that eq. (\ref{421}) is valid for $k>0$, namely for 
the transition between two expanding branches with ${\rm sign} [H]={\rm 
sign}  
[\dot\b]={\rm sign} [k] >0$, and that it 
implies $0<R_k<1$ ($R_k \ra 0$ for $k\ra 
\infty$, as expected when the effective potential in eq. (\ref{418}) 
becomes negligible). If we consider transitions between contracting 
branches, $k<0$, the probability becomes $e^{2\pi k}$ so that, in 
general, $R_k=e^{-2\pi|k|}$. The appearance of an always negative 
argument in the exponential is a general consequence of applying 
tunneling boundary conditions in superspace, as clearly stressed 
recently also in \cite{23}.

\vskip 2 cm
\renewcommand{\theequation}{5.\arabic{equation}}
\setcounter{equation}{0}
\section{Conclusion}

In string cosmology, a classical description of the background evolution 
based on the low energy string effective action is allowed both at early 
and late times (i.e. at large time scales in string units), but it is 
not allowed in the intermediate epoch, when the background is expected to 
exit from the inflationary regime. A classical model of transition 
from the initial string perturbative vacuum to 
the present standard cosmological regime conflicts both with 
phenomenological constraints and with formal no-go theorems. 

In this paper we have shown that such a transition can be studied 
quantum mechanically, and can be formulated as a problem of reflection 
of the Wheeler-De Witt wave function in superspace. By using tunneling 
boundary conditions, we find that the transition can occur (with finite 
probability) even in the case of background configurations in which it 
would be classically forbidden.

This quantum approach to the graceful exit problem is free from 
ambiguities of operator ordering, because of the underlying $O(d,d)$ 
symmetry of the kinetic part of the Hamiltonian. It thus seems to 
provide an appropriate tool for a systematic classification of the 
initial conditions, irrespective of the (unknown) kinematic details of 
the high curvature, strong coupling, transition regime.

Various problems, in this approach, are still to be solved, such as the 
physical interpretation of the wave function and the univocal choice of 
an appropriate time parameter (these problems affect in general the WDW 
approach to quantum cosmology, not only our scenario). We believe, 
however, that the result presented in this paper may improve our 
understanding of the birth and of the evolution of the Universe in terms 
of the basic principles of string theory.

\vskip 2 cm 

\section*{Acknowledgements}

This work was supported in part by the ``Human Capital and Mobility 
Program" of 
the European Commission, under the CEE contract No. ERBCHRX-CT94-0488.


\newpage


\begin{thebibliography}{99}

\bibitem{1}M. Gasperini and G. Veneziano, Astropart. Phys. 1 (1993) 317 

\bibitem{2}M. Gasperini and G. Veneziano, Mod. Phys. Lett. A8 (1993) 
3701

\bibitem{3}M. Gasperini and G. Veneziano, Phys. Rev. D50 (1994) 2519

\bibitem{4}R. Brustein and G. Veneziano,
Phys. Lett. B329 (1994) 429 

\bibitem{5}N. Kaloper, R. Madden and K. A. Olive, Nucl. Phys. B452 
(1995) 677

\bibitem{6}N. Kaloper, R. Madden and K. A. Olive, {\it Axions and the 
graceful exit problem in string cosmology}, UMN-TH-1414/95 
(hep-th/9510117)

\bibitem{7}R. Easther, K. Maeda and D. Wands, {\it Tree-level string 
cosmology}, SUSSEX-AST-95/9-1 (hep-th/9509074)

\bibitem{8}E. Kiritsis and C. Kounnas, Phys. Lett. B 331 (1994) 51

\bibitem{9}M. C. Bento and O. Bertolami, Class. Quantum Grav. 12 (1995) 
1919

\bibitem{10}J. E. Lidsey, Phys. Rev. D52 (1995) R5407

\bibitem{12}C. Lovelace, Phys. Lett. B135 (1984) 75;

E. S. Fradkin and A. A Tseytlin, Nucl. Phys. B261 (1985) 1;

C. G. Callan et al., Nucl. Phys. B262 (1985) 593.

\bibitem{13}M. Gasperini, in {\it Proc. of the 2nd Journ\'ee Cosmologie}, 
ed. by N. Sanchez and 

H. De Vega (World Scientific, Singapore, 1995) p.429

\bibitem{14}K. A. Meissner and G. Veneziano, Mod. Phys. Lett. A6 (1991) 
3397;  Phys. Lett. B 267 (1991) 33.

\bibitem{15}M. Gasperini and G. Veneziano, Phys. Lett. B 277 (1992) 256

\bibitem{16}G. Veneziano, Phys. Lett. B265 (1991) 287

\bibitem{17}A. A. Tseytlin, Mod. Phys. Lett. A6 (1991) 1721

\bibitem{18}B. S. De Witt, Phys. Rev. 160 (1967) 1113

\bibitem{19}A. Ashtekar and R. Geroch, Rep. Prog. Phys. 37 (1974) 1211

\bibitem{19a}M. Muller, Nucl. Phys. B337 (1990) 37

\bibitem{20}C. Angelantonj, L. Amendola, M. Litterio and F. Occhionero, 
Phys. Rev. D51 (1995) 1607

\bibitem{21}M. Abramowicz and I. A. Stegun, Handbook of Mathematical 
Functions (Dover, New York, 1972)

\bibitem{22}A. Vilenkin, Phys. Rev. D33 (1986) 3650; 
Phys. Rev. D37 (1988) 888. 

\bibitem{23}A. Vilenkin, {\it Predictions from quantum cosmology} 
(gr-qc/9507018), to appear  
in  Proc. of the Int. School of Astrophysics 
``D. Chalonge" (Erice, September 1995), ed. by N. Sanchez (Kluwer Acad. 
Publ., Dordrecht, 1995) 


\end{thebibliography}
\end{document}